\documentclass{emulateapj}
\usepackage{lscape,afterpage}
\slugcomment{Astronomical Journal, Accepted 2005 April 13th}

\shorttitle{GALEX Ultraviolet Variability Catalog}
\shortauthors{WELSH et al.}

\begin{document}

%% LaTeX will automatically break titles if they run longer than
%% one line. However, you may use \\ to force a line break if
%% you desire.

\title{THE {\it GALEX}  ULTRAVIOLET VARIABILITY (GUVV) CATALOG}

%% Use \author, \affil, and the \and command to format
%% author and affiliation information.
%% Note that \email has replaced the old \authoremail command
%% from AASTeX v4.0. You can use \email to mark an email address
%% anywhere in the paper, not just in the front matter.
%% As in the title, you can use \\ to force line breaks.

\author{
Barry Y. Welsh,\altaffilmark{1}
Jonathan M. Wheatley,\altaffilmark{1}
Kenneth Heafield,\altaffilmark{2}
Mark Seibert,\altaffilmark{2}
Stanley E. Browne,\altaffilmark{1}
Samir Salim,\altaffilmark{3}
R. Michael Rich,\altaffilmark{3}
Tom A. Barlow,\altaffilmark{2}
Luciana Bianchi,\altaffilmark{4}
Yong-Ik Byun,\altaffilmark{5}
Jose Donas,\altaffilmark{6}
Karl Forster,\altaffilmark{2}
Peter G. Friedman,\altaffilmark{2}
Timothy M. Heckman,\altaffilmark{4}
Patrick N. Jelinsky,\altaffilmark{1}
Young-Wook Lee,\altaffilmark{5}
Barry F. Madore,\altaffilmark{7}
Roger F. Malina,\altaffilmark{6}
D. Christopher Martin,\altaffilmark{2}
Bruno Milliard,\altaffilmark{6}
Patrick Morrissey,\altaffilmark{2}
Susan G. Neff,\altaffilmark{8}
David Schiminovich,\altaffilmark{2}
Oswald H. W. Siegmund,\altaffilmark{1}
Todd Small,\altaffilmark{2}
Alex S. Szalay,\altaffilmark{4} and
Ted K. Wyder\altaffilmark{2} }

\altaffiltext{1}{Experimental Astrophysics Group, Space Sciences Laboratory, University of California, 7 Gauss Way, Berkeley, CA 94720; bwelsh@ssl.berkeley.edu, wheat@ssl.berkeley.edu}

\altaffiltext{2}{California Institute of Technology, MC 405-47, 1200 East
California Boulevard, Pasadena, CA 91125}

\altaffiltext{3}{Department of Physics and Astronomy, University of
California, Los Angeles, CA 90095}

\altaffiltext{4}{Center for Astrophysical Sciences, The Johns Hopkins
University, 3400 N. Charles St., Baltimore, MD 21218}

\altaffiltext{5}{Center for Space Astrophysics, Yonsei University, Seoul
120-749, Korea}

\altaffiltext{6}{Laboratoire d'Astrophysique de Marseille, BP 8, Traverse
du Siphon, 13376 Marseille Cedex 12, France}

\altaffiltext{7}{Observatories of the Carnegie Institution of Washington,
813 Santa Barbara St., Pasadena, CA 91101}

\altaffiltext{8}{Laboratory for Astronomy and Solar Physics, NASA Goddard
Space Flight Center, Greenbelt, MD 20771}

%% Notice that each of these authors has alternate affiliations, which
%% are identified by the \altaffilmark after each name.  Specify alternate
%% affiliation information with \altaffiltext, with one command per each
%% affiliation.

%% Mark off your abstract in the ``abstract'' environment. In the manuscript
%% style, abstract will output a Received/Accepted line after the
%% title and affiliation information. No date will appear since the author
%% does not have this information. The dates will be filled in by the
%% editorial office after submission.

\begin{abstract}
 We present Version 1.0 of the NASA
 Galaxy Evolution Explorer ($\it GALEX $) ultraviolet variability catalog (GUVV) that
 contains information on
84 time-variable and transient sources gained with simultaneous near and far ultraviolet
photometric observations.
These time-variable sources were serendipitously revealed 
in the various 1.2$^{\circ}$ diameter star fields currently being
surveyed by the $\it GALEX$ satellite in
 two ultraviolet bands (NUV 1750~-~2750~\AA, FUV 1350~-~1750~\AA) with limiting
AB~magnitudes of 23 - 25. The largest-amplitude variable
objects presently detected by $\it GALEX$ are M-dwarf flare stars,
which can brighten by 5~-~10~mag in both
the NUV and FUV bands during short duration ($<$ 500 s) outbursts.
Other types of large-amplitude ultraviolet
variable objects include $ab$-type RR~Lyrae stars,
which can vary periodically by 2~-~5~mag in the $\it GALEX$ FUV band. This
first GUVV catalog
lists galactic positions and possible source identifications  in
order to provide the astronomical community with a list of time-variable objects
that can now be repeatedly observed at other wavelengths.
We expect the total number of time-variable source detections 
to  increase as
the $\it GALEX$ mission progresses, such
that later version numbers
of the GUVV catalog will contain substantially more variable sources.

\end{abstract}

%% Keywords should appear after the \end{abstract} command. The uncommented
%% example has been keyed in ApJ style. See the instructions to authors
%% for the journal to which you are submitting your paper to determine
%% what keyword punctuation is appropriate.


%% From the front matter, we move on to the body of the paper.
%% In the first two sections, notice the use of the natbib \citep
%% and \citet commands to identify citations.  The citations are
%% tied to the reference list via symbolic KEYs. The KEY corresponds
%% to the KEY in the \bibitem in the reference list below. We have
%% chosen the first three characters of the first author's name plus
%% the last two numeral of the year of publication as our KEY for
%% each reference.

\section{Introduction}

The primary scientific mission of the NASA Galaxy Evolution Explorer
($\it GALEX$) satellite \citep{mar05} is to
explore star formation processes and the
histories of galaxies through imaging photometric observations
in two ultraviolet bands (NUV 1750 - 2750\AA\, FUV 1350 - 1750\AA).
However, $\it GALEX$ is also making
serendipitous ultraviolet photometric measurements 
of several million
stars and other non-galactic objects during the course of its All-Sky 
Imaging Survey (AIS) and during its deeper repeated observations of selected
small areas
of the sky with its Deep Imaging Survey (DIS) and
Medium Imaging Survey (MIS). In particular,
$\it GALEX$ has a high sensitivity, low background noise, a wide field of view ($1.^{\circ}2$),
and it makes repeated visits to deep fields \citep{mor05}. These observational
capabilities have enabled
the detection of numerous variable and transient ultraviolet sources, 
many of which exhibit much larger amplitudes of variation in the
ultraviolet region than that recorded at visible wavelengths.

One good example of a serendipitous source-detection by $\it GALEX$ is the
RR Lyrae star, ROTSE-I J143753.84+345924.8 \citep{wheat05}. Using a series
of 38 separate $\it GALEX$ pointings a 4.9 AB magnitude
variation was observed in the FUV band, compared with only a 0.8 magnitude
variation at visible wavelengths. From these ultraviolet  light-curves it was possible
to constrain theoretical models that placed meaningful limits on both the temperature
and metallicity of the star. One further example of a $\it GALEX$ serendipitous observation
is that of the massive ultraviolet flare on the dM4e star, GJ 3685A, in which an overall
brightness increase of AB $>$ 4 magnitudes was observed in both the FUV and
NUV bands in a time
period of only 60 seconds \citep{rob05}. Other types of astronomical source that $\it GALEX$
can potentially detect are cataclysmic variables, Cepheid variables, soft X-ray
transients and (possibly) gamma-ray bursters.

In this Paper we list 84 variable and transient ultraviolet sources that have been detected during
the first 15 months of the $\it GALEX$ all-sky survey, which is
currently envisaged to be completed within the next 18 months. These present
observations (taken from data covering $\sim$ 10$\%$ of the sky) will form the basis of an
increasing data-base of variable UV sources whose physical properties can be further explored
in more detail by the astronomical community in other wavelength bands.

\section{Observations and Data Analysis}
We have used the $\it GALEX$ FUV and NUV-band photometric imaging data
recorded during the period
June 2003 to August 2004, which reside in the Multi-Misison
Archive at the Space Telescope Science Institute (MAST).
During this time-period the $\it GALEX$ satellite performed
several types of imaging and spectroscopic observations using its 1.2$^{\circ}$ field
of view. We have restricted our analysis to data recorded in the
photometric imaging mode by the two
FUV and NUV photon-counting detectors \citep{jelin03}. These imaging observations
consist of data recorded in three observational modes, each with different exposure
times. They consist of (i) the All-Sky Imaging Survey (AIS) which observes 
regions of the sky for $\sim$ 100 seconds. Adjacent AIS sky-fields have a small (2$\%$) 
area of overlap that enables a limited number of detections of 
source variability between consecutive survey images, (ii) the Medium Imaging Survey
(MIS), which observes regions of the sky with a total exposure of
 $\sim$ 1500 seconds (i.e. one $\it GALEX$ orbit), that in some cases
have been repeatedly observed in order to gain a better resultant
S/N ratio of the sky-field, and (iii) the
Deep Imaging Survey (DIS) which repeatedly observes specific
pre-selected regions of interest on the sky for
many ($>$ 20) orbits in order to
accumulate a total exposure time of $\sim$ 30,000 seconds for each selected sky-field. 
In addition to these 3 main types of survey mode, $\it GALEX$ is also
carrying out a survey of bright nearby galaxies (NGS), which entails recording
both photometric imaging and spectroscopic data on selected fields for
periods typically of $\sim$ 2000 seconds. 
In Table 1 we list the total area of the sky (in square degrees) observed thus far by the various
$\it GALEX$ surveys, together with the respective number of variable
 sources detected. Not surprisingly, the
largest number of variable source detections per unit area have been found in
the Deep Imaging Survey mode (i.e. about one source per field).  

\begin{table*}  %%[htbp] 
\caption{\label{obslog_1} GALEX survey information}
%%\begin{flushleft}
\begin{tabular}{lccc}
\hline
\hline
Survey  & Square Degrees & Number of Variable & Detections per \\
Mode&Surveyed& Sources Detected & Square Degree\\
\hline
AIS&2729&52&0.02\\
MIS&129&13&0.10 \\
DIS&15&18&1.2 \\
NGS& 49& 8 & 0.16\\
\hline
\hline
\end{tabular}
%%\end{flushleft}
\end{table*}

All of these currently observed sky-fields are located
well away from the galactic plane in order to avoid saturation
of the detectors due to overly-bright stellar sources and
to avoid regions where interstellar absorption is high.
The recorded data files contain photon events that
have been processed by the standard $\it GALEX$ Data Analysis Pipeline operated
at the Caltech Science Operations Center (Pasadena, CA) that ingests time-tagged
photon lists, instrument and spacecraft housekeeping data and
satellite pointing aspect information \citep{mor05}. The
data pipeline uses a source detection
algorithm (called Sextractor) to produce a catalog of 
source positions and corresponding ultraviolet magnitudes for
each observation. Comparison software was then run on this
catalog to detect sources that we deem as being
either variable or transient.
Variable sources are defined as being present in repeatedly observed 
fields and exhibiting an orbital variation
greater than AB = 0.3 magnitudes  (with an associated
change $>$ 3$\sigma$ in the magnitude error)
in their derived FUV and/or NUV magnitudes recorded in two or
more separate observations.
Transient sources are defined as objects which are detected
only once, in the FUV and/or NUV bands, in a repeatedly
observed star-field. We note that a `transient source' may
be a variable star detected only once near maximum light.

In Table 2 we list 84 sources that have
been identified as being either variable (V) or transient (T) events by
the aforementioned $\it GALEX$ ultraviolet sky-field observations. 
We have not included several asteroids that
appeared as potential transient UV objects in the data.
The catalog number of each of the
detected sources is listed in column (1)
of this Table. This number is a unique identifier containing
both the Right Ascension (J-2000.0) of the source in hours,
minutes and
decimal seconds 
and the
corresponding source Declination (J-2000.0)
in degrees, minutes and decimal seconds. These
positions are typically accurate to $\pm$ 1.0 arc sec for sources that
have been observed in the central 1$^{\circ}$ of the detectors \citep{mor05}.
In column (2) we list whether the source is variable (V) or transient (T), based
on the citerion listed in the previous paragraph.
Column (3) lists the USNO-B1.0 all-sky catalog
designation  \citep{mon03}, where available, for 
the source that is closest (and within $\pm5$ arc seconds) to 
the position of the object listed in column (1). 
In column (4) we provide a possible identification for the source based 
on objects listed in the Simbad on-line astronomical catalog for
targets with positions that are coincident within $\pm5$ arc sec
of the $\it GALEX$ determined position,
and in column (5) we list the most likely type of
astronomical source-type for that object. Criteria used to
make this latter determination are generally varied, but (for the
brighter sources) are mainly based
on either their Simbad catalog identifications or on
inspection of their $\it GALEX$ UV light-curve data.
Flare stars were found to be generally bright just once during a series of UV observations,
whereas
periodic variables (PVs) exhibited a large range of values
in their measured UV magnitudes. Some of these PVs are listed
as RR Lyrae stars in Simbad and such designations have been
used accordingly in column (5).

In Column (6) we list the $\it GALEX$ survey mode (AIS, MIS, DIS or NGS)
of the sky field in which
the object was discovered. Columns (7) and (8) respectively list the total number of
observations of the particular sky field in the NUV
channel and the number of these exposures in which
the source was detected. Column (9) lists the
maximum observed NUV magnitude for the source (measured
over one AIS, MIS, DIS or NGS integration period) and column (10)
lists the variation between the correspondingly measured maximum
and minimum NUV magnitudes (i.e. $\Delta$m).
Similarly, columns (11) - (14) list the equivalent number of observations, number
of detections, maximum magnitude and variation in magnitude for the FUV channel.
Here we note that the non-detection of
a source previously observed in both (or one) of the two UV-bands
can be attributed to either intrinsic variability (i.e. an astrophysical effect) or by
the fact that one of the detectors was turned off
during a particular observation for instrument safety reasons.
Finally, columns (15) and (16) list
the respective $\it g$ and $\it r$ magnitudes as recorded by the
Sloan Digital Sky Survey (SDSS) catalog \citep{abaz03} for the source designation
listed in column (3). Stars with uncertain SSDS magnitudes (due to detector
saturation and other effects) are marked with an asterisk (*).
We also note that as the data pipeline software matures and
refines over the
extent of the $\it GALEX$ mission, the derived NUV and FUV source magnitude
values may alter slightly. It is hoped that later versions of
the GUVV catalog, based on the entire $\it GALEX$ data archive,
will be forthcoming.

\section{Discussion}
The 84 $\it GALEX$ variable and
transient sources listed in Table 2 have been observed using
$\sim$ 2600 separate observations that cover a total of
$\sim$ 3000 sq. degrees of sky. Variable sources (72) account for
86$\%$ of the listed sources and the remaining 12 sources are all of a transient nature.
Only one of the sources, J090054.7+303113.3, was detected as being
simultaneously transient
in $\it both$ of the FUV and NUV channels. The remaining transient sources
were singularly detected in only one of the two $\it GALEX$ channels,
presumably due to the different sensitivity limits of each UV
photometric band.

Several of the GUVV sources
are previously known flare stars, 
RR Lyrae stars, quasars or X-ray sources, but the large majority of the 
UV variable objects in this catalog have no previously listed source
identification. 
We note that \cite{sieb05} have produced color-color diagrams for
$>$ 350,000 (non-variable) objects detected by $\it GALEX$ in a
143 sq. degree portion
of the sky that overlaps with that of the Sloan Digital Sky Survey (SDSS) \citep{abaz03}.
Plots of ($\it g$ - $\it r$ ) versus (m$_{nuv}$ - $\it g$) magnitude have
revealed a segregation of main sequence, horizontal branch,
white dwarf, sub-dwarf and M dwarf stellar populations in
the $\it GALEX$ data. It has been
found that the number 
density of M dwarf - white dwarf binary systems is at least
twice as high in
the $\it GALEX$ NUV data than that found using the SSDS $\it u$-magnitude.
In order to assess if there are any similar underlying
fundamental observable parameters that may classify these 84 detections into
distinct variable source sub-groups we have produced the following Figures
in which the data have been plotted with 3 different symbols identified
with known flares stars (open circles), known RR Lyrae stars (filled
circles) and  
and sources with no identification (crosses).
    
In Figure 1 we plot
values of ($\it g$ - $\it r$ ) versus
(m$_{fuv}$ - $\it g$) magnitude and
see that the vast majority of
the data points divide into two separate groupings
with (i) 0.4 $>$ ($\it g$ - $\it r$) $>$ -0.2 and
(ii) 1.6 $>$ ($\it g$ - $\it r$) $>$ 1.0. The former data
group (in the lower region of Figure 1)
contains all of the identified RR Lyrae stars in
the $\it GUVV$ catalog and
the latter data grouping (in the upper region of Figure 1)
contains two dMe flare stars.
With regard to the former data group we note that 
 \cite{ivezic05} have found very similar 
color-color limits for RR Lyrae
stars based solely on their SSDS ($\it g$ - $\it r$)
magnitudes. Since all of the data points contained in Figure 1 are
targets that have been deemed to be variable based on their
$\it GALEX$ UV observations,
then the un-identified sources in the lower region
of this
figure are most probably also RR Lyrae (or $\delta$ Sct) type stars.
We are currently obtaining low resolution visible spectra of the
un-identified sources in the upper region of Figure 1 to determine if
they are dMe flare stars. If the stars in this regime do turn out
to be dMe flares then plots like Figure 1 may represent a very
useful
tool for the future selection
of previously unidentified RR Lyrae and dMe flare
stars based solely on their $\it GALEX$ UV variability and
their SDSS ($\it g$ - $\it r$) magnitudes.

\begin{figure}
{\includegraphics[height=8cm]{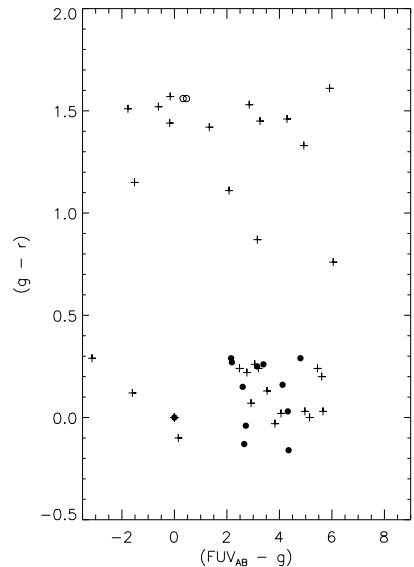}}
\caption{Plot of the SDSS ($\it g$ - $\it r$) magnitudes versus (m$_{fuv}$ - $\it g$) magnitudes for the GUVV catalog sources. Note the division of the sources into two distinct groups with different ($\it g$ - $\it r$) magnitudes. The lower grouping of targets contains mainly  RR Lyrae stars (filled circles) and the upper grouping are probably mostly dMe flare stars (open circles). The remainder of the targets are un-identified GUVV sources (crosses).} 
\label{Figure 1}
\end{figure}

\begin{figure}
{\includegraphics[height=8cm]{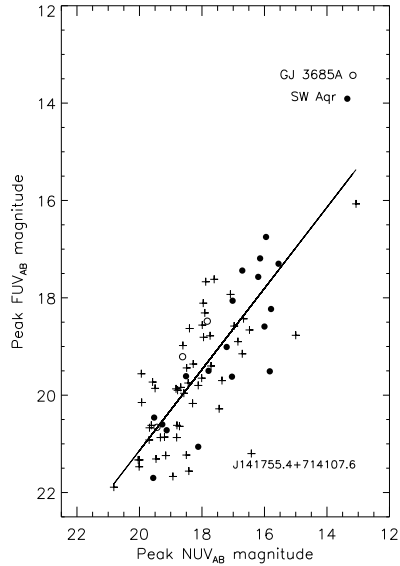}}
\caption{Plot of the $\it GALEX$ peak FUV  AB magnitude versus the
peak NUV AB magnitude for the GUVV catalog sources. The majority of the targets lie within $\pm$ 1 magnitude of a straight-line of slope +0.81. The 3 exceptions are highlighted on the figure. See Figure 1 text for an explanantion of the plotting symbols. } 
\label{Figure 2}
\end{figure}
    
In Figure 2 we plot the observed $\it GALEX$ peak FUV magnitude versus the
peak NUV magnitude. We see that the majority of these sources lie within $\sim$
$\pm1.0$ magnitudes of the best-fit straight-line (of slope +0.81).
However, three
sources, J211517.8+000432.5 (the RR Lyare star, SW Aqr),
J114740.7+001521.0 ( the dM4 flare star, GJ 3685A ) and
J141755.4+714107.6
lie well outside of these limits. Apart from the loose
proportionality between the FUV and NUV peak magnitudes for all these 
sources, we see no other underlying physical discriminators in this figure. 

\subsection{Interesting GUVV objects}
Of the 26 GUVV sources that possess reliable identifications
in column (4) of Table 2, we note that 14 are known RR Lyrae
variables, 4 are previously known dMe flare stars,
and 5 are radio and/or X-ray sources. The vast majority
of the GUVV sources have no
firm identification and therefore are clearly prime targets for follow-up 
ground-based photometric and spectroscopic observations. To highlight some of
the interesting astronomical sources that
have thus far been identified in the GUVV
catalog, in the following two sections we illustrate some of the scientific studies that can
be explored using these new UV data.

\subsubsection{RR Lyrae Stars}
$\it GALEX$ is well-suited for the detection
of RR Lyrae variable stars since they
are typically found to
vary by 2 - 6 magnitudes in the FUV band, with a corresponding
NUV magnitude change of $\sim$ 50$\%$ of this value. This can be compared with
a typical change in magnitude
of only $\sim$ 1.0 recorded at visible wavelengths \citep{skil93}.
In Figure 3 we show the $\it GALEX$ NUV and FUV light-curves for
the star GUVV-J100133.3+014328.4 that show AB magnitude variations of
$\sim$ 4.9 in the FUV band and $\sim$ 2.0 in the NUV band \citep{browne05}.
These curves were constructed using a least-string software program
that derived a period of 0.543 days for the light-curve from
82 NUV and 26 FUV observations.
It is clear from these UV variability data that this star is of the RR Lyrae
type (parenthetically we note that its $\it g$ - $\it r$ value
of 0.03 magnitudes places it in the lower region of Figure 1 together with the
other RR Lyrae stars). 
Since the derived FUV flux from
$\it GALEX$ observations of this class of variable star is
highly sensitive to a change in the Kurucz model atmosphere temperature of
only $\sim$ 100 - 200K, these UV data taken together with visible
light-curve observations can enable better estimates to be made of the
stellar metallicity.
However, although a large flux variation can be observed from
RR Lyrae stars in the FUV (and NUV), the current
sensitivity limit
of the $\it GALEX$ observations for these faint sources presently
limits their
potential use as (non-galactic) cosmic distance scale indicators.
Thus, although $\it GALEX$ may well prove to be a rich source of
galactic RR Lyrae star detection,
their stellar distances are probably better determined 
using follow-up visible observations with
large aperture ground-based telescopes.

\begin{figure}
\center
{\includegraphics[height=8cm]{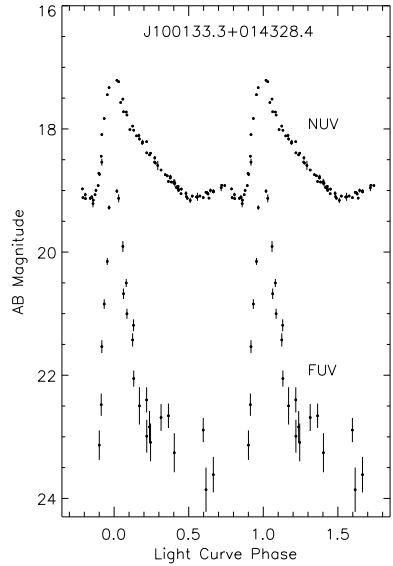}}
\caption{ Observed $\it GALEX$ FUV and NUV light-curves for the star GUVV-J100133.3+014328.4, which we identify as a new RR Lyrae star. The phase was computed using a derived period of 0.543 days.  } 
\label{Figure 3}
\end{figure}

\subsubsection{Flare Stars}
$\it GALEX$ has been fortunate
to detect many large short-lived outbursts of UV flux, typically
lasting $<$ 200 seconds, generated by dMe-type flare stars. This emission
is linked to magnetic processes occurring in their outer
stellar atmospheres (coronae).
In one extreme case of
the dM4e star GJ 3685 (GUVV-J114740.7+001521.0), its overall UV brightness 
increased by more than a factor of 10,000 making it 20 times larger than any
previously observed
UV flare \citep{rob05}. It should be noted that the changes in the
FUV and NUV
magnitudes listed in columns (10) and (14) for these events are average
values recorded
over one $\it GALEX$ observation period. Only through inspection of
the individual time-tagged photon events, recorded
with a time resolution of 0.005 sec for each of these observational
periods, can the underlying physical properties
of the flare-mechanism be
revealed. In the case of the previously mentioned UV flare, the
$\it GALEX$ observations have 
detected two major outburst events separated by 200 seconds
that were accompanied by numerous short-duration ($<$ 10 sec) micro-flares
during the entire 1600
second observation. In Figure 4 we show the NUV light curve
(gained from time-tagged photon data) for the flare
recorded by $\it GALEX$ on
the star GUVV-J144738.47+035312.1 on June 3rd 2004, that shows a more
modest brightness increase that consists of
at least 4 major outbursts observed during a $\sim$ 150 second flaring interval.
We note that
although the brightest flaring M dwarf stars detected by $\it GALEX$ have
distances typically $<$ 30 pc and
can inject up to $\sim$ 10$^{34}$ erg into the surrounding interstellar gas,
based on the number of these events recorded
thus far by $\it GALEX$ this UV flux
is insufficient to make a significant contribution to the global
ionization properties of the local interstellar medium.

\begin{figure}
\center
{\includegraphics[height=8cm]{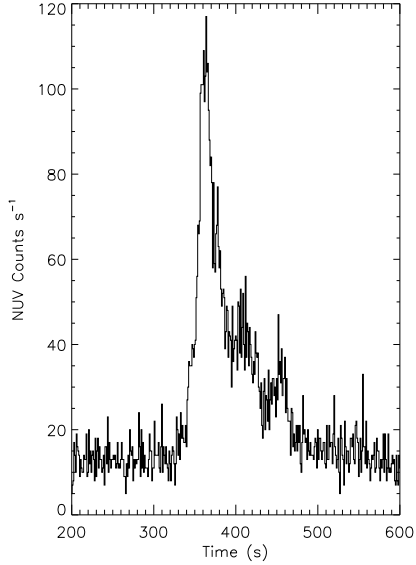}}
\caption{Near UV emission as a function of time for the flare recorded on the star GUVV-J144738.47+035312. Note at least two subsidiary emission events that followed within 100 seconds of the main flare. } 
\label{Figure 4}
\end{figure}

\begin{acknowledgments}
We gratefully acknowledge NASA's support for construction, operation,
and science analysis for the GALEX mission,
developed in cooperation with the Centre National d'Etudes Spatiales
of France and the Korean Ministry of
Science and Technology. We acknowledge the dedicated
team of engineers, technicians, and administrative staff from JPL/Caltech,
Orbital Sciences Corporation, University
of California, Berkeley, Laboratoire d'Astrophysique de Marseille,
and the other institutions who made this mission possible.
Financial support for this research was provided by NASA grant
NAS5-98034. This publication makes use of data products from the SIMBAD database,
operated at CDS, Strasbourg, France.
\end{acknowledgments}

\pagestyle{empty}
\afterpage{
\begin{landscape}
{
\renewcommand{\arraystretch}{1.0}
\begin{table*}[htbp]
\caption{\label{obslog_2}$\it GALEX$ Ultraviolet Variability Catalog, V1.0.}
\begin{tiny}
\begin{flushleft}
\begin{tabular}{lccllcrrccrrcccc}
\hline
\hline
&&&&&Discovery&NUV  &&&&FUV&&&&SDSS DR3&\\
GUVV&&USNO-B1.0&ID&Type&Survey&$N_{obs}$&$N_{det}$&Max&$\Delta m$&$N_{obs}$&$N_{det}$&Max&$\Delta m$&g&r\\
\hline
J004347.9+421654.9&V&1322-0015866&CC And&Delta Sct variable&NGS&9&9&13.02&0.63&9&9&16.07&0.93&-&-\\
J004548.2-435509.1 &V&0460-0006826&&PV  &DIS&10&10&16.00&1.38&10&9&18.59&3.77&-&-\\
J010732.6+360956.5&V&1261-0017670   &1RXS J010732.1+361001&x-ray source&NGS&2&2&20.01&0.60&2&2&21.47&0.55&-&-\\
J085218.1+311047.2 &V&1211-0164505   && &AIS&3&3&15.00&0.33&3&3&18.77&0.37&15.23*&15.19*\\
J090054.7+303113.3&T&-&&&AIS&2&1&19.69&-&2&1&20.92&-&22.44&21.29\\
J090808.2-004610.9 &V&0892-0180755&&&MIS&3&3&17.99&2.47&2&2&18.56&2.88&15.40&14.53\\
J090904.4+091714.4 &T& - &&&AIS&3&1&17.96&-&2&1&18.11&-&22.33&22.08\\
J091324.0+091417.9 &T&0992-0184047   && &AIS&3&1&19.93&-&2&1&20.15&-&17.30&15.77\\
J092458.8+021834.1 &T&0923-0226013   && &AIS&2&1&19.31&-&2&0&-&-&10.83*&10.50*\\
J092551.9+015545.6 &T&0228-01607-1   &HD 81463(?)&A0 star&AIS&2&1&17.36&-&2&1&19.70&-&11.79*&14.30*\\
J092620.4+034541.8 &V&0937-0186908   &&&AIS&2&2&17.91&1.88&2&2&18.31&1.95&19.91&19.79\\
J092851.8+041630.0 &V&0942-0172841   &FIRST J092851.8+041630&Radio source&AIS&2&2&17.95&1.13&2&2&18.81&0.37&18.66&18.76\\
J093026.0+071221.6 &V&0972-0216018   &WW Leo&RR Lyrae-ab&AIS&2&2&15.83&0.97&2&1&19.51&-&12.28*&13.95*\\
J095801.1+021250.0 &V&0922-0237968   && &DIS&93&2&19.68&3.42&81&1&20.67&-&18.59&17.48\\
J095816.1+014843.6 &T&0918-0208424   && &DIS&47&1&20.04&-&42&1&21.33&-&21.49&19.92\\
J100133.3+014328.4 &V&0917-0193609   &&PV&DIS&96&82&17.21&2.00&87&26&19.01&4.85&14.69&14.66\\
J100141.5+020758.8 &V&0921-0232170   && &DIS&140&97&20.82&2.22&126&10&21.89&1.80&17.60&16.14\\
J100152.1+021158.5 &V&0921-0232199   && &DIS&93&89&18.80&3.10&84&70&20.62&2.58&13.97&15.86*\\
J100209.5+020726.5 &V&0921-0232253   && &DIS&93&2&19.58&2.38&84&2&19.73&2.65&21.50&19.99\\
J100358.9-270001.4 &V&0629-0308953   && &NGS&5&5&18.26&1.43&3&0&-&-&-&-\\
J102002.7+611538.9 &V&1512-0181964   && &AIS&4&4&16.71&1.62&3&2&19.15&0.85&14.00&14.00\\
J102525.7-392130.8 &V&0506-0220494   &&&NGS&6&6&18.72&1.95&5&2&20.64&0.20&-&-\\
J104844.1+581539.4 &V&1482-0239659   &&PV &DIS&48&47&18.51&2.88&24&6&19.61&2.45&17.01&16.86\\
J105513.7+564747.0 &V&1467-0226668   &&PV& DIS&65&62&17.04&2.27&26&9&19.62&2.94&15.50&15.34\\
J105622.2+570520.6 &V&1470-0242660   &&&DIS&62&62&17.02&2.33&26&9&18.06&4.49&15.87&15.60\\
J105926.1-005927.9 &V&0890-0199535&SDSS J105926.11-005927.6 &RR Lyrae-ab&DIS&3&3&19.53&2.09&3&1&20.46&-&18.30&18.01\\
J111147.3+510549.4 &V&1410-0219003  && &AIS&4&4&18.12&1.50&3&1&21.06&-&16.26&15.97\\
J112334.9+474014.6&V&1376-0264908&&&AIS&4&4&19.45&1.07&3&0&-&-&17.30&17.05\\
J113340.3+502328.0 &V&1403-0230850   &CZ UMa&RR Lyrae-ab&AIS&4&4&16.71&2.93&2&1&17.44&-&14.78&14.91\\
J114740.7+001521.0 &V&0902-0204368   &GJ 3685A&dM4e flare&MIS&4&4&13.17&6.67&2&2&13.43&7.69&14.37*&12.84*\\
J120157.2-183153.7 &V&0714-0246553   &&&NGS&5&4&18.83&1.92&5&1&19.87&-&-&-\\
J122034.6-030947.7 &V&0868-0272366   &&&AIS&2&2&19.77&0.51&2&0&-&-&15.05&13.98\\
J122057.1+673838.9 &V&1576-0166067   &1RXS J122057.4+673845&x-ray source&MIS&3&3&16.67&1.96&2&2&18.43&3.06&13.29*&11.31*\\
J122415.6-014914.0 &V&0881-0268821   &[VZA2004] 195&RR Lyrae&AIS&2&2&19.27&1.34&1&1&20.60&-&16.25&16.41\\
J122743.3-005754.4 &V&0890-0214215&[VZA2004] 199&RR Lyrae&MIS&4&4&19.13&1.82&1&1&20.72&-&17.57&17.32\\
J122836.9-064230.0 &V&0832-0270055   && &AIS&2&2&19.62&2.39&1&1&20.62&-&-&-\\
J123313.6+020029.1 &V&0920-0259726   &&&NGS&2&2&19.33&1.92&2&1&20.87&-&17.67&17.43\\
 J123349.3-024456.2 &V&0872-0322120   && &MIS&3&3&19.16&1.87&3&1&21.24&-&16.27&16.24\\
J123512.5+621744.6 &T&1522-0247069   && &DIS&72&1&18.61&-&41&1&18.98&-&19.59&18.07\\
J123738.0-040841.2 &V&0858-0232491   && &AIS&2&2&18.50&0.94&2&1&21.23&-&-&-\\
J123913.5-113307.4 &V&0784-0251397   &NSVS 123913-113314&Radio source&NGS&2&2&17.09&2.33&2&1&17.93&-&-&-\\
J124109.1+230159.8 &T&1130-0231256   && &AIS&2&1&18.81&-&1&1&20.87&-&-&-\\
J124328.2-055431.7 &T& - && &AIS&2&1&19.94&-&1&1&19.56&-&-&-\\
J124746.2+243940.6 &V&1146-0199030   && &AIS&6&5&18.58&2.25&4&2&19.96&0.77&-&-\\
%%\hline
%%\multicolumn{10}{l}{* = uncertain SSDS magnitude, PV = periodic variable } \\
%%\hline
%%\hline
%%\end{tabular}
%%\end{flushleft}
%%\end{tiny}
%%\end{table*}
%%}
%%\end{landscape}

%%\clearpage
%%}
%%\newpage

%%\afterpage{
%%\begin{landscape}
%%{
%%\renewcommand{\arraystretch}{1.0}
%%\begin{table*}[htbp]
%\caption{\label{obslog_3}$\it GALEX$ Ultraviolet Variability Catalog (continued)}
%%\begin{tiny}
%%\begin{flushleft}
%%\begin{tabular}{lccllcrrccrrcccc}
%%\hline
%%\hline
%%&&&&&Discovery&NUV  &&&&FUV&&&&SDSS DR3&\\
%%GUVV&&USNO-B1.0&ID&Type&Survey&$N_{obs}$&$N_{det}$&Max&$\Delta m$&$N_{obs}$&$N_{det}$&Max&$\Delta m$&g&r\\
%%\hline
J124812.4+005737.4 &V&0909-0215691   &FASTT 524 &variable star&AIS&2&2&18.93&0.95&2&1&21.67&-&16.06&15.86\\
J124906.9-010421.9 &V&0889-0220289   &BW Vir   &RR Lyrae-ab&MIS&4&3&15.95&2.54&4&2&16.75&0.39&14.03&14.07\\
J125000.6+310824.0 &V&1211-0200416   &TX Com&RR Lyrae-ab&AIS&4&4&16.2&1.97&1&1&17.57&-&-&-\\
J125409.7+252707.8 &V&1154-0198874   &EN Com&RR Lyrae-ab&AIS&6&4&18.79&1.55&0&0&-&-&-&-\\
J125428.5+003739.5 &V&0291-00256-1   && &AIS&2&2&16.24&2.61&0&0&-&-&11.91*&11.39*\\
J125905.7+242632.9 &V&1144-0198736   && &AIS&3&3&17.61&3.67&2&2&17.62&3.04&-&-\\
J125911.1+263745.1 &V&1166-0214036   &Ton 682&UV-excess object&AIS&6&4&17.74&0.15&4&4&18.78&1.25&-&-\\
J130204.5+463533.7 &V&1365-0231982   && &AIS&2&2&18.01&1.45&2&1&19.65&-&15.82&15.85\\
J130213.6+241420.0 &V&1142-0198264   &BF Com&RR Lyrae-ab&AIS&2&2&15.55&1.04&2&2&17.3&4.17&-&-\\
J130615.1+293657.5 &V&1196-0208146   &EV Com&RR Lyrae-ab&DIS&8&8&19.56&1.91&8&2&21.7&1.62&-&-\\
J130934.8+285905.9 &V&1189-0208615   &GJ 1167A&dM5 flare&DIS&6&5&19.44&2.07&3&3&20.66&1.53&-&-\\
J131012.3+474517.0 &V&1377-0296991   &1RXS J131011.9+474521  &&AIS&2&2&19.47&1.24&2&1&21.31&-&15.40&13.79\\
J131855.8+433100.0 &V&1335-0237412   && &AIS&2&2&18.42&1.33&2&1&21.56&-&16.11&15.87\\
J132135.3+431145.3 &V&1331-0281572   && &AIS&4&3&18.28&2.27&4&1&19.36&-&16.60&16.38\\
J132546.6-425140.9 &V&0471-0361002   && &NGS&5&5&18.49&1.78&5&5&19.44&1.58&-&-\\
J132715.2+425932.1 &V&1329-0296231   && &AIS&4&3&17.88&3.63&4&1&17.67&-&20.81&20.52\\
J133052.6-031644.6 &T&0867-0280752   && &AIS&2&1&18.63&-&0&0&-&-&22.11&21.02\\
J133057.0-040824.6 &V&0858-0243273&& &AIS&4&4&16.44&0.56&0&0&-&-&-&-\\
J133115.8+405657.6 &V&1309-0239624   && &AIS&2&2&16.97&1.41&4&1&18.58&-&14.52&14.50\\
J133757.1+401610.8 &V&1302-0233853&&&AIS&4&4&20.00&0.91&4&1&21.33&-&18.41&18.34\\
J134156.2+030744.3 &V&0931-0264765   && &MIS&3&3&17.45&1.41&3&2&20.28&1.04&14.62&14.59\\
J135408.2+573615.9 &V&1476-0290232   && &AIS&2&2&18.76&1.25&0&0&-&-&15.09&14.74\\
J140113.3+710524.0 &V&1610-0100824   && &AIS&2&2&16.48&1.91&2&1&18.66&-&-&-\\
J141755.4+714107.6 &V&4406-00241-1   && &AIS&2&2&16.42&0.70&2&1&21.20&-&-&-\\
J142329.4+034317.6 &V&0937-0238977   && &MIS&4&2&19.20&2.96&3&1&20.86&-&15.93&14.60\\
J142551.2+042949.3 &V&0944-0225286   &&UV flare&MIS&7&3&18.62&4.54&5&1&19.21&-&18.88&17.32\\
J143741.1+344119.0 &T&1246-0218818   && &DIS&84&1&18.40&-&84&1&18.63&-&18.81&17.37\\
J143753.7+345923.9 &V&1249-0218617   &ROTSE1 J143753.84+345924.8&RR Lyrae-ab&DIS&84&84&15.79&2.78&84&75&18.23&4.85&13.78*&13.49*\\
J144433.4+364200.2 &V&1267-0242808   && &AIS&3&2&17.71&1.71&2&2&19.40&0.99&16.14&14.69\\
J144708.1+345158.4 &T&1248-0221943   && &AIS&6&1&19.50&-&4&1&19.86&-&18.53&17.11\\
J144738.4+035311.8 &V&0938-0240178   &&UV flare&MIS&3&3&17.83&4.24&3&1&18.48&-&18.02&16.46\\
J145110.2+310639.7 &V&1211-0222759&1RXS J145110.3+310648&x-ray source&DIS&22&21&18.68&0.88&11&10&19.84&0.97&-&-\\
J145339.2+501151.6 &V&1401-0265491   && &AIS&2&2&18.12&1.48&2&1&19.80&-&16.74&16.48\\
J150957.7+621334.9 &V&1522-0274109   && &AIS&2&2&18.30&1.97&2&1&20.17&-&14.12&13.36\\
J151120.0+392036.5 &V&1293-0251929   && &AIS&3&3&16.85&1.22&2&1&18.90&-&14.71&15.38*\\
J151234.9+392416.1 &V&1294-0251325   && &AIS&3&3&18.44&1.94&2&1&19.75&-&16.22&16.09\\
J151532.1+364806.4 &V&1268-0255264   && &AIS&3&3&18.78&2.57&2&1&19.90&-&17.42&17.18\\
J164940.0+345820.3&V&1249-0245815&HL Her&RR Lyrae-ab&DIS&4&4&17.79&1.76&4&2&19.50&2.59&16.11&15.85\\
J203853.9-580358.6&V&0319-1042594&UU Ind&RR Lyrae-ab&MIS&2&2&16.14&2.50&2&1&17.19&-&-&-\\
J211517.8+000432.5&V&0900-0581544&SW Aqr&RR Lyrae-ab&MIS&3&3&13.35&2.19&3&3&13.91&6.32&11.77*&11.49*\\
\hline
\multicolumn{10}{l}{* = uncertain SSDS magnitude, PV = periodic variable} \\
\hline
\hline
\end{tabular}
\end{flushleft}
\end{tiny}
\end{table*}
}
\end{landscape}
}

\end{document}